# Localized pore evolution assisted densification during spark plasma sintering of nanocrystalline W-5wt.%Mo alloy

Ajeet K. Srivastav[1,*], Niraj Chawake[2], Devinder Yadav[3] , N.S. Karthiselva[4], B.S. Murty[4]

[1]*Department of Metallurgical and Materials Engineering,*

*Visvesvaraya National Institute of Technology, Nagpur – 440 010, India*

[2]*Erich Schmid Institute of Materials Science, Jahnstraße 12, 8700 Leoben, Austria*

[3]*Department of Metallurgical and Materials Engineering,*

*Indian Institute of Technology Patna – 801 103, India*

[4]*Department of Metallurgical and Materials Engineering,*

*Indian Institute of Technology Madras, Chennai – 600 036, India*

**Abstract:**

The present work reports the role of different atomic mobility induced localized pore evolution on densification during spark plasma sintering of nanocrystalline W-Mo alloy powder. The shrinkage (or expansion) behavior of cold compacted milled powders was studied using dilatometry during non-isothermal sintering up to 1600 °C. Subsequently, the milled powders were densified to ~95% relative density using spark plasma sintering up to 1600 °C. The enhanced localized Joule heating due to dynamically evolved porous structure could be attributed for the densification during spark plasma sintering.

Spark plasma sintering (SPS) has shown a great promise over the last few decades due to several advantages than the conventional sintering processes, such as densifying variety of materials to fully dense structure at a lower temperature in a short duration of time with a possibility of retaining the fine grain structure [1–3]. These benefits are primarily attributed to localized heating mechanism [4,5]. However, localized heating mechanisms are highly dependent on the nature of contact between the particles in due course of sintering process of conductive materials. It has been established that the localized heating and electrosintering mechanisms will be highest and largely contributing at the initial stages of sintering (with more point contacts and higher current) and gradually decreases with time [6,7]. The nature of particle contacts during the sintering could be entirely different in presence of another elemental powder particles than the pure metallic powders. Further, the varying thermal induced mobility responses of atoms of various elements poses a challenge to consolidate a multi-component system [8,9]. It has been shown that the Kirkendall effect might play an important role in the evolution of particle morphologies and contacts therein in addition to the densification kinetics [10,11]. Moreover, it is expected that the electric current assisted heating and densification mechanisms will be operative for the longer time period with the evolving Kirkendall pores during the SPS of multi-components systems.

---

* Corresponding author; email addresses:srivastav.ajeet.kumar@gmail.com; ajeet.srivastav@mme.vnit.ac.in

The present paper reports the role of Kirkendall porosity on densification process during both non-isothermal sintering and the SPS of nanocrystalline W-5wt.%Mo alloy. The study shows that such systems could be densified using SPS which is otherwise difficult to sinter by conventional processes.

Mechanical milling/alloying (MM/MA) of W and W-5wt.%Mo were performed using high-energy planetary ball mill (Fritsch Pulversitte-5, Germany). WC balls of 10 mm diameter in tungsten carbide vials were used as grinding medium. Toluene was used as a process control agent. Milling was performed at 300 rpm for 6 h with a ball to powder weight ratio of 5:1.

The X-ray diffraction measurements were carried out using X'Pert Pro (PANalytical, The Netherlands) X-ray diffractometer with Cu-$K_\alpha$ radiation in a range of 10–140 degree using 0.02° step size with a 30 s time per step. Double-Voigt analysis [12] was espoused to assess the crystallite size and its distribution after fitting the XRD patterns using a pseudo-Voigt function in HighScore Plus (version 3.0e). The adopted methodology for line profile analysis has been discussed in an earlier report [13,14]. Transmission electron microscopy was performed on milled nanocrystalline powders using Philips CM12 operated at 120 kV.

Non-isothermal sintering was performed on cold compacted 8 mm diameter cylindrical pellet of W and W-5wt.%Mo alloy using Dilatometer (Setaram, Setsys Evolution) in Argon atmosphere with a constant heating rate of 20 K min$^{-1}$ and up to 1600 °C under ~0.6 N load. The shrinkage data was used to understand the effect of Mo alloying on densification behavior of nanocrystalline W during the non-isothermal sintering.

The consolidation of milled/alloyed powders was performed using SPS (Dr Sinter SPS-650 machine, Sumitomo Metals, Japan). The densification of milled/alloyed powders was performed in a cylindrical graphite die of 10 mm diameter up to 1600 °C by heating at 100 °C min$^{-1}$ with a dwell time period of 5 min and 50 MPa uniaxial pressure applied throughout the process. The density of sintered samples was measured using Archimedes' principle using water as a medium.

The non-isothermally sintered samples were investigated using a scanning electron microscope (FEI QUANTA 400 manufactured by FEI, USA). Electron backscattered diffraction (EBSD) was carried out on the electropolished samples in a FEG SEM (Inspect F, FEI, USA) equipped with TSL-OIM software using a step size of 300 nm.

Nanoindentation study of SPS samples following the Oliver and Pharr method [15] was carried out to quantify the nanohardness and elastic modulus using Hysitron Triboindenter (TI 950, Hysitron, USA) with a Berkovich indenter at a maximum load of 8000 μN.

The XRD patterns of milled W and W-5wt.%Mo powders in comparison to unmilled W powder are illustrated in Fig. 1(a). The magnified (211) peak as shown in Fig. 1(b) clearly indicates the peak broadening due to crystallite size refinement and induced lattice strain during milling [16,17]. Complete alloying could not be ascertained just by observing the peak profile due to low solute content and almost matching peak positions for both W and Mo [13]. The average crystallite size and its distribution is depicted in Fig. 1(c). The average crystallite size and its distribution for W-5wt.%Mo is 44 ± 3 nm after MM/MA. The nanocrystalline nature of the milled powder was confirmed using the TEM micrograph (Fig.1d) and the corresponding diffraction pattern for nanocrystalline W powder.

The starting W and Mo powder particle sizes lies in the range of 3-5 μm with nearly spherical in shape as shown in Fig.S1 (supplementary material) . However, the morphology of the milled W and W-5wt.%Mo was observed as typical flaky due to the very nature of the mechanical milling process. The bigger dimension of the milled W-5wt.%Mo was almost similar with the starting powder size as illustrated in Fig. S2 (supplementary material). The non-isothermal sintering shrinkage behavior of nanocrystalline powders is compared with the unmilled W powder in Fig. 2(a). There was an expansion for W-5wt.%Mo sample, unlike nanocrystalline W and unmilled W powders. This could be attributed to Kirkendall porosity evolved during the sintering process of W-5wt.%Mo sample. However, it is difficult to distinguish here the Kirkendall porosity and the remnant porosity after certain extent of densification. The existence of Kirkendall porosity could be supported with two inferences here. First, it has been shown earlier that W and Mo show Kirkendall porosity either as diffusion couple or during sintering [18–20]. Secondly, to understand formation of Kirkendall pores, we performed dilatometry experiment in the regime where the starting sample shows only expansion and no densification. Therefore, we sintered W-5wt.%Mo non-isothermally up to 1450°C as the expansion was observed approximately up to 1450°C as shown in the Fig.2(a). The Fig. 2(a) inset shows the representative SEM micrograph as the porous structure evolved due to differential diffusion kinetics of W and Mo resulting in Kirkendall porosity in W-5wt.%Mo non-isothermally sintered up to 1450°C. The SEM micrographs are shown in the Fig.S3 (supplementary material). The densely appeared flakes in milled W-5wt.%Mo as illustrated in Fig.S2 transform to the porous structure as shown in Fig.S3 (supplementary material) and this could be the reason behind overall expansion of the sample as appeared in the dilatometry data. The average pore size was estimated as $0.88 \pm 0.20$ µm as shown in the supplementary material (Fig. S4). It is worth mentioning here that the microstructure becomes rich in point contacts developed in due course of dynamic pore evolution during the non-isothermal sintering of the W-5wt.%Mo alloy sample.

However, the densification curves during SPS of nanocrystalline W and W-5wt.%Mo powders demonstrate the typical *S*-curve as illustrated in Fig. 2(b). The relative density for both the samples was observed almost equal to ~95%. Furthermore, the densification kinetics is slower for W-5wt.%Mo alloy in comparison to W during most of the consolidation period. This could be ascribed to the competition between dynamically evolved Kirkendall porosity and the thermally induced diffusion driven densification process. It should be noted here that the densification starts during the non-isothermal sintering and the faster densification kinetics during SPS of W-5wt.%Mo sample was observed at almost the similar temperatures (~1450 °C). The densification behavior and its competition with dynamic pore evolution due to Kirkendall effect during SPS of W-5wt.%Mo alloy are discussed later.

The EBSD characterization was carried out for spark plasma sintered samples. The inverse pole figure maps and the analyzed average grain size and grain size distributions are shown in Fig. 3. The significant grain growth during SPS of both W and W-Mo alloy could be explained with higher fraction of random high angle grain boundaries (RHAGBs) in ball milled nanocrystalline powders [21]. The RHAGBs have more tendency for thermal migration than the low energy boundaries and therefore more vulnerable to coarsening. The average grain size is lower for W-5wt.%Mo sample (7.3 µm) in comparison to W (11.4 µm). The elemental distribution as shown in the supplementary materials (Fig. S5) illustrates the homogeneous distribution of Mo with no indication of any preferential segregation. This shows that the solute-drag to restrict the grain boundary mobility could be the reason behind lower grain sizes for the W-5wt.%Mo alloy sample.

The nanoindentation study was performed to evaluate the mechanical properties of the SPS W and W-5wt.%Mo samples. The representative load and penetration depth curves are illustrated as Fig. S6 in the supplementary material. The elastic modulus and hardness values were compared in Fig. S7 (supplementary material). The ascertained hardness $6.1\pm0.8$ GPa and elastic modulus $296\pm31$ GPa of the consolidated W-5wt.%Mo alloy sample was slightly less than that of W $7.2\pm0.2$ GPa, $350\pm10$ GPa, respectively. Further, we will be focusing on to understand the densification mechanism during SPS of W-Mo alloy.

In general, the densification (pore closure) is massive atomic movement caused by temperature, deformation or both. Therefore, a high temperature or high pressure or both are required for the densification to occur. The electric current, temperature and stress contributed multi-faceted densification mechanisms operating during SPS are different and difficult to understand than those during conventional sintering [1]. The peculiar heating mechanisms such as Joule heating and formation of spark and plasma are responsible for the faster densification achieved in the case of SPS [22]. Joule heating is prominent heating mechanisms for the metal/alloy powders and occurs at the particle-particle contacts due to the resistance provided to the current flowing through them [5]. Thus, it is dependent on the density or porosity and nature of contacts between the powder particles. More the porosity, the higher is the probability of point contacts and so the Joule heating locations. Also, the temperature experienced at particle-particle contacts (localized temperature) may be much higher than the actual measured temperature [23]. Recently, it has been suggested that the applied current enhances the mass transport and dislocation mobility and thereby improving the plastic flow of the material during SPS [24]. It is worthy to point out here that the sintering of multi-component system prefers to reaction/homogenization over densification due to larger chemical driving forces than the capillary forces [25]. As a result, a void formation may take place.

The above discussion has direct implications for understanding the densification behavior of W and W-5wt.%Mo powders. In the present study, it has been established through non-isothermal sintering using dilatometry (Fig. 2a) that W-5wt.%Mo shows the formation of Kirkendall porosity during sintering, while W doesn't show this phenomenon (Fig. 2a). The different atomic mobility results in higher vacancy concentration for the faster diffusing species and subsequently, the supersaturation of vacancies leads to the formation of Kirkendall voids/porosity [26]. In addition, the possibility of such pores could be increased in presence of pressure even with the equilibrium concentration of vacancies [27]. It is reported earlier that the diffusion phenomenon gets enhanced under the pulsed electric current which is typical to SPS process [28]. In the presence of an electric field, the intrinsic diffusivities of elements can be significantly enhanced by orders of magnitude [29]. Thus, the formation of Kirkendall pores in W-5wt.%Mo during SPS can act as a potential site for Joule heating by increasing the localized resistance to the pulsed electric current. Thereby, a large population of Kirkendall pores induced higher localized temperature is possible in W-5wt.%Mo as compared to W. Also, during SPS the spatial temperature distribution results in improved diffusion coefficients [22,29,30]. Recently, Deng *et al* [24] have shown that the diffusion coefficient governs the densification of the powders. Thus, if we consider the Kirkendall pore of a diameter ($R$), then the temperature gradient ($\nabla$T) around it can be estimated using the following formula [31,32]:

$$\nabla T \approx \frac{1}{R} \sqrt{\frac{\sigma_0 T_0 E_0^2 \Delta \tau}{2 C_m \eta}} \qquad (1)$$

where, $\sigma_0$-electrical conductivity, $T_0$-initial temperature, $E_0$-intensity of an electric field, $\Delta\tau$-time of an electric field, $C_m$-the specific heat of the material, $\eta$-number of electric impulses. The $\nabla T$ estimated for a micron size Kirkendall pore formed in W for 12:3 ms electric pulse at 1400 °C with $E_0 \approx 5$ V/5 mm (obtained from the SPS experiment) is approximately 1064 K/μm i.e. ≈ 800 °C/μm.

The temperature gradient around the pores generates the vacancy gradient ($\nabla C_v$) and the vacancy flux ($J$) is given by the following equation [31,32]:

$$J = D_v \left[ \frac{k_T}{T \nabla T} - \nabla C_v \right] \qquad (2)$$

where, $k_T$-thermal diffusivity and $D_v$-diffusion coefficient of vacancies.

From Equation (2), it can be concluded that the vacancy diffusion will occur from the large to small pores leading to shrinkage of large pores as illustrated in Fig. S11 (supplementary material) [33].

Additionally, the vacancy mediated massive unequal atomic mobility of both W and Mo during SPS leads to the annihilation of excess vacancies at sinks such as dislocations and grain boundaries. Such events could be enhanced for the starting milled powders having nanocrystalline structure and rich in defect density, in particular, the dislocation density. The annihilation of excess vacancies becomes the reason for dislocation sources and their mobility and thereby contributing towards material flow [34]. Therefore, it is believed that such vacancy annihilation phenomenon imparts plastic flow of the material during SPS.

Thus, the scenario during sintering of W-5wt.%Mo system is more complex. There are two competing phenomena occurring, first one is the formation of Kirkendall pores increasing localized electric resistance and generate large thermal gradient ($\nabla T \approx 800$ °C/µm). This leads to the second phenomenon that is the shrinkage of large pores (Equation 2). In due process, the annihilation of excess vacancies aids to the plastic flow. However, it's challenging to bifurcate contributions of these phenomena. Also, simultaneous application of pressure makes it difficult to separate out these effects. The localized intensification of pressure with a higher localized temperature at the pore surface can finally result in pore closure too, thus improving the densification. It is known in the SPS literature that the simultaneous application of pressure enhances densification [1].

Intriguingly, the formation of Kirkendall pores will increase the resistance offered to the electric current and thus to support this discussion, the resistance is calculated from densification data (Current and Voltage) obtained during SPS for W-5wt.%Mo and W powders [5]. The Fig. 4(a) shows the variation of resistance with temperature during SPS taking into consideration the instantaneous porosity (refer supplement for calculations). It is clear that W-5wt.%Mo system offers higher resistance as compared to W thus causing more Joule heat and leading to its densification. The resistance trajectory could be divided into four sub-parts. First, resistance increases slightly and becomes constant followed by a sudden decrease indicating the densification phenomena and saturates finally at the lowest value. First and second parts are typically common for both W and W-5wt.%Mo. The W-5wt.%Mo resistance trajectory saturates almost to the same value but at a temperature nearly 200 °C higher as compared to W. This could be well understood from the Fig. 4(b) where the variation of resistance with temperature is plotted with the relative density for the W-5wt.%Mo. The initial increase of resistance could be attributed to thermal effects as the density is almost constant. Subsequently, the thermal effects compete with initial stages of densification and the resistance becomes almost constant. These two stages are almost the same for both W and W-5wt.%Mo. In the third stage, the sudden linear drop is an indication of densification where the pore closure dominates over thermal effects for the resistance variation [7]. Finally, the resistance gets saturated with the densification curve.

The above observation has its significance for systems which show Kirkendall porosity during sintering. The Kirkendall porosity is the hindrance for achieving full density during conventional sintering. However, the localized heating mechanisms in SPS can be advantageous to achieve higher density by turning these pores as potential heating sites. In addition, the plastic flow of the material is believed to be enhanced due to the annihilation of vacancies so created which aids up with other material flow mechanisms such as electromigration during SPS [35].

In conclusion, the role of Kirkendall porosity during the sintering processes of nanocrystalline W-5wt.%Mo alloy was studied. The Kirkendall porosity induced expansion was observed during the non-isothermal sintering of W-5wt.%Mo alloy. However, the typical *S*-curve densification behavior with ~95% relative density was evidenced during the SPS. This could be explained with enhanced localized heating due to dynamic pore evolution induced increased point contacts. The study clearly shows that the systems showing the Kirkendall porosity could be consolidated using SPS.

**Acknowledgements:**

Ajeet K. Srivastav acknowledges the financial support from CSIR-India (via award no. 09/084(0519)2010-EMR-I).

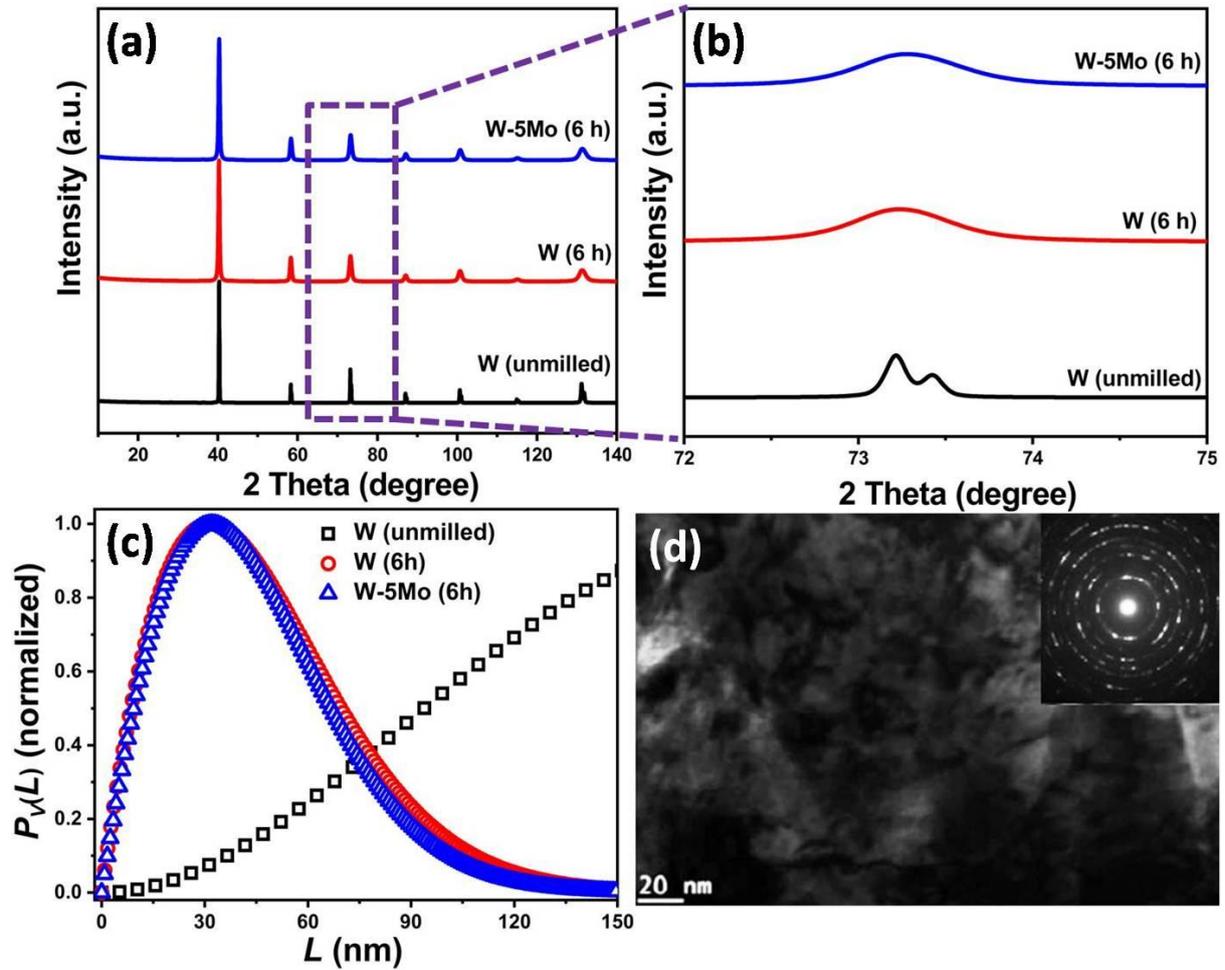

**Fig. 1:** (a) XRD patterns of milled W and W-5wt.% Mo powder showing broadened peak profiles in comparison to the unmilled W, and (b) magnified (211) peak indicates the alloying after the milling process, and (c) effect of milling on average crystallite size distribution using double-Voigt methodology, and (d) TEM micrograph of milled W powder corroborating the nanograin structure and size distribution as observed in XRD analysis.

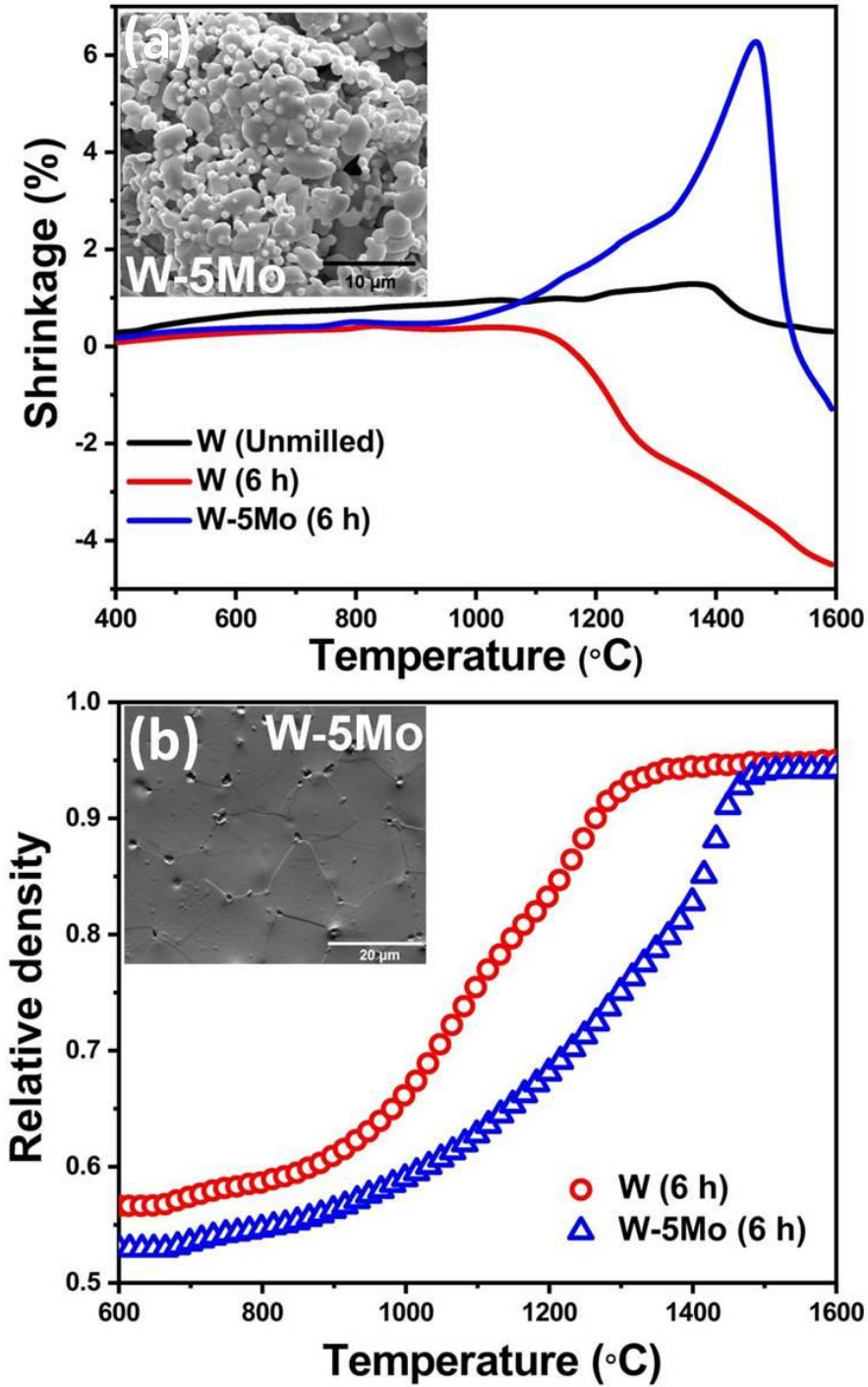

**Fig. 2:** (a) Kirkendall porosity (inset) induced expansion during non-isothermal sintering of W-5wt.% Mo in contrast to milled and unmilled W powder, and (b) relative density versus temperature showing the dense SPS W-5wt.% Mo alloy (inset) in comparison to non-isothermal sintering. The SEM in the inset shows the final microstructure of W-5wt.% Mo after (a) non-isothermal sintering up to 1450°C, and (b) SPS up to 1600°C.

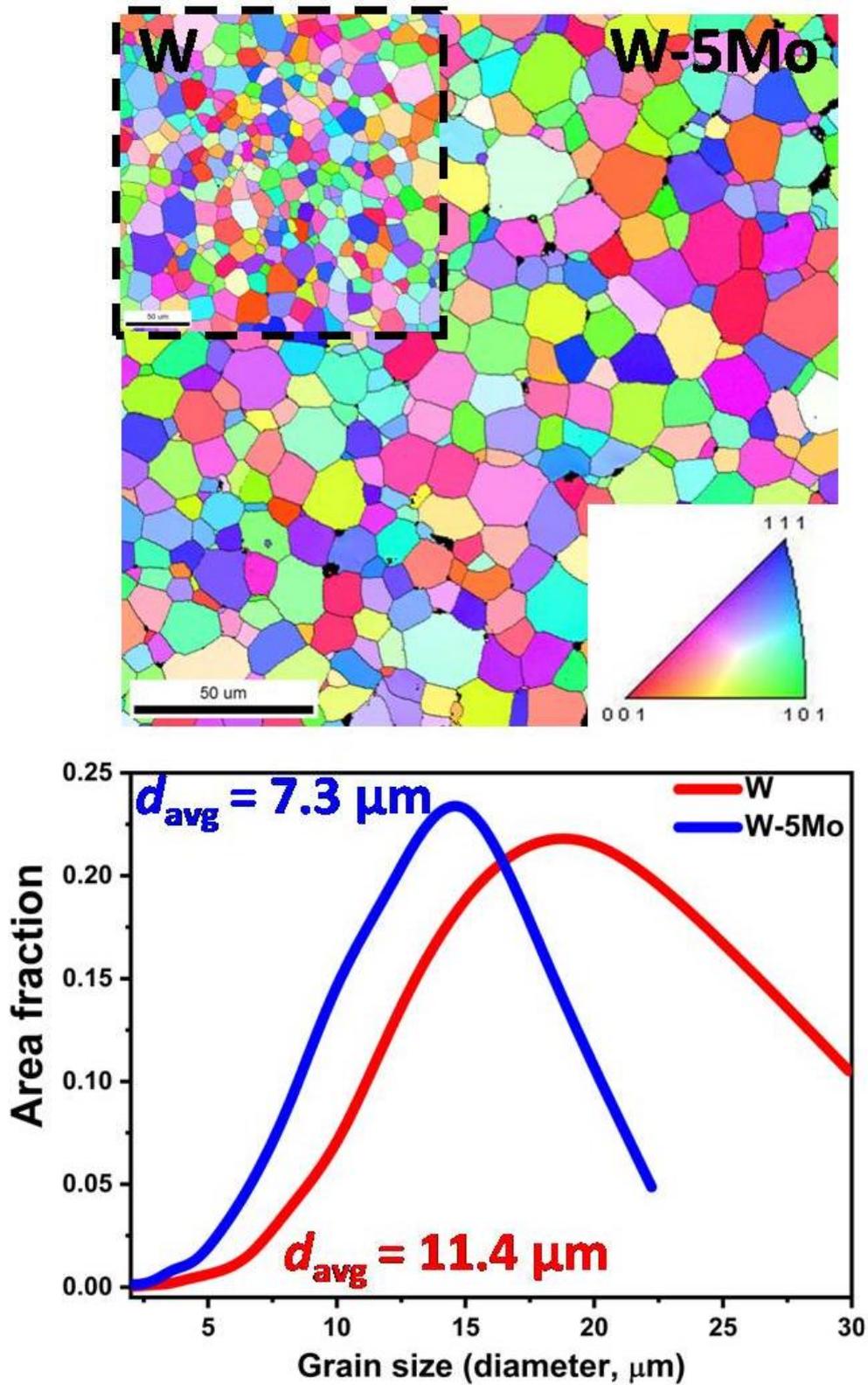

**Fig. 3:** EBSD showing grain orientation mapping of SPS W-5wt.% Mo alloy. The grain size distribution as analyzed using EBSD data indicates smaller average grain size and narrower size distribution for W-5wt.% Mo alloy in comparison to W.

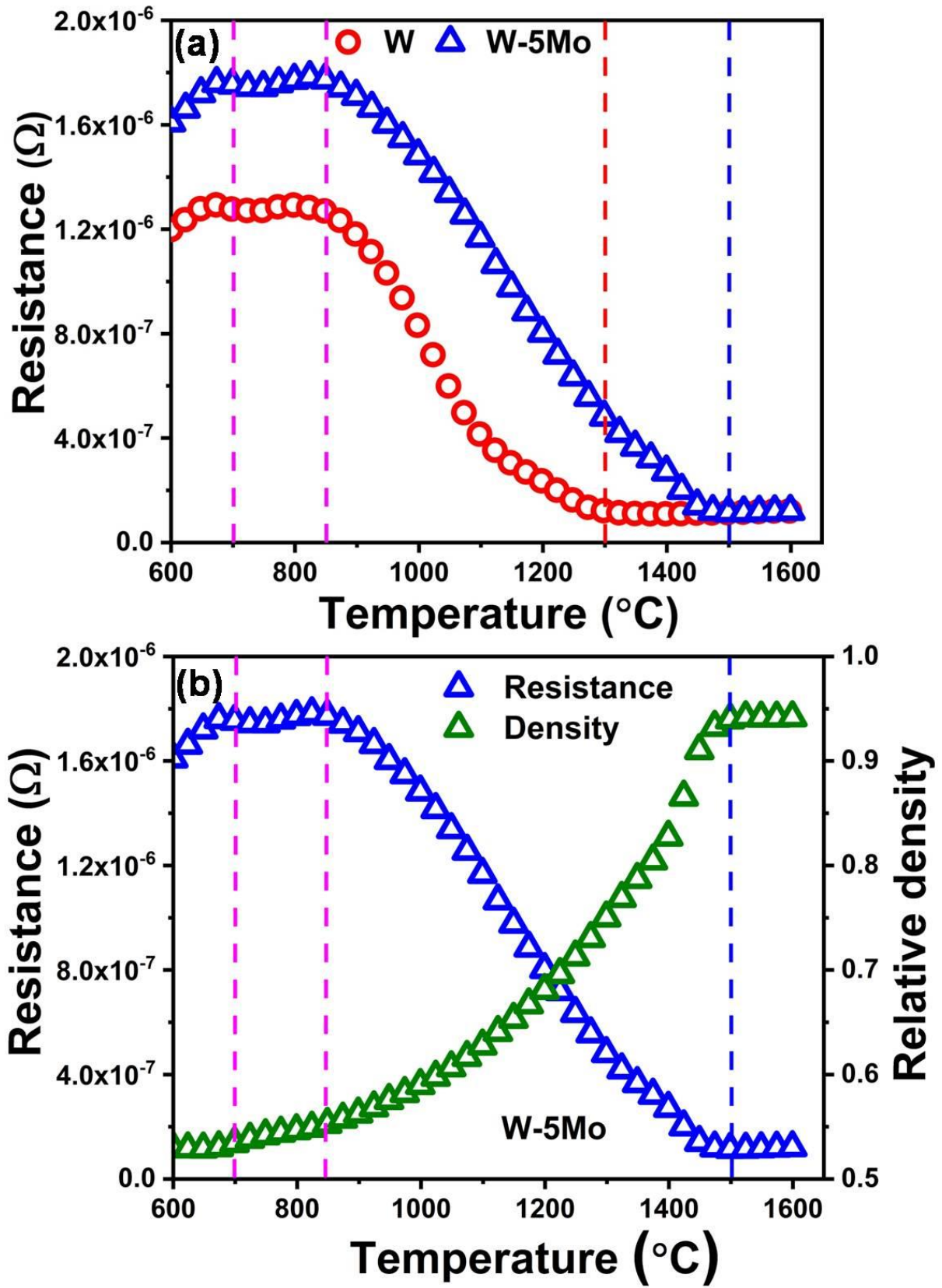

**Fig. 4:** (a) Resistance variation with temperature of W-5wt.% Mo alloy in comparison to W, (b) resistance and relative density variation with temperature during SPS of W-5wt.% Mo.